\documentclass[
               DIV=15, font size=10.5pt,
                 onecolumn,
               ]{scrartcl}  

\usepackage{color} 
\usepackage{psfrag} 
\usepackage{subfig}
\usepackage{hyperref}
\usepackage{amsmath}
\usepackage{graphics}
\usepackage{mathtools}
 \usepackage{pstool} 
\usepackage{todonotes}
\usepackage{gensymb}
\usepackage{bm}
\usepackage{multirow}

\usepackage{textcomp}
          
\usepackage[square, authoryear]{natbib}
 
\usepackage{pgfplots}
\usepackage{pgfplotstable}
\usepackage{filecontents}

\usepackage{tikz}
\usepackage{tikzscale}

\usetikzlibrary{spy}
\usetikzlibrary{snakes,arrows,shapes}    
\usetikzlibrary{spy}
\usetikzlibrary{backgrounds}  
\usetikzlibrary{decorations}
\usetikzlibrary{positioning}
\usetikzlibrary{patterns}
\usetikzlibrary{calc} 

\usepackage{booktabs}   
\usepackage{makecell} 
\usepackage{colortbl}   

\usepackage{xspace}
\usepackage{color}     
\usepackage{setspace}   
 
\usepackage{soul}
\usepackage{ulem}
\usepackage{currfile}
\usepackage{import}

\usepackage{authoraftertitle} 
\usepackage{authblk} 
\usepackage{cleveref}

\pgfplotsset{compat=1.8}

\newcommand{\findmax}[3]{
    \pgfplotstablesort[sort key={#2},sort cmp={float >}]{\sorted}{#1}%
    \pgfplotstablegetelem{0}{#2}\of{\sorted}%
    \let #3=\pgfplotsretval%
}

\pgfplotscreateplotcyclelist{myCycleListBlackAndWhite}
{%
  {black, mark=o},
  {black, mark=square},
  {black, mark=triangle},
  {black, mark=diamond}, 
  {black, mark=pentagon}, 
  {black, mark=*},
  {black, mark=square*},
  {black, mark=triangle*},
  {black, mark=diamond*}, 
  {black, mark=pentagon*}, 
}

\definecolor{darkgreen}{rgb}{0,0.4,0} 
\definecolor{darkbrown}{rgb}{0.5, 0.396, 0.09}

\definecolor{c1}{rgb}{0.0, 0.4196078431372549, 0.6431372549019608}
\definecolor{c2}{rgb}{1.0, 0.5019607843137255, 0.054901960784313725}
\definecolor{c3}{rgb}{0.6705882352941176, 0.6705882352941176,
0.6705882352941176} \definecolor{c}{rgb}{0.34901960784313724, 0.34901960784313724, 0.34901960784313724}
\definecolor{c4}{rgb}{0.37254901960784315, 0.6196078431372549,
0.8196078431372549} \definecolor{c}{rgb}{0.7843137254901961, 0.3215686274509804, 0.0}
\definecolor{c5}{rgb}{0.5372549019607843, 0.5372549019607843,
0.5372549019607843} \definecolor{c}{rgb}{0.6352941176470588, 0.7843137254901961, 0.9254901960784314}
\definecolor{c6}{rgb}{1.0, 0.7372549019607844, 0.4745098039215686}
\definecolor{c7}{rgb}{0.8117647058823529, 0.8117647058823529,
0.8117647058823529}

\pgfplotscreateplotcyclelist{myCycleListColor}
{%
  {blue, mark=o},
  {red, mark=square},
  {darkbrown, mark=triangle},
  {darkgreen, mark=diamond},
  {black, mark=pentagon},
  {blue, mark=*},
  {red, mark=square*},
  {darkbrown, mark=triangle*},
  {darkgreen, mark=diamond*},
  {black, mark=pentagon*},
}

\pgfplotsset{every axis/.append style= 
              {
                font=\small,
                mark size=2,
                line width = 0.1,
                legend style={font=\small, mark size=3, draw=none, fill=none},
                legend cell align=left,
                cycle list name=myCycleListColor,
              }
            }
            
\pgfplotstableset
{
  every odd row/.style={before row={\rowcolor[gray]{0.9}}},
  every head row/.style=
  {
    before row=
    {%
      \toprule
    },
    after row=\midrule
  },
  every last row/.style=
  {
    after row=\bottomrule
  }
}

\newif\ifdrawboundingbox

\usetikzlibrary{external} 
\tikzset{external/system call={pdflatex \tikzexternalcheckshellescape
-halt-on-error -interaction=batchmode -jobname "\image" "\texsource"}} 

\pgfkeys
{ 
  /pgf/number format/.cd,
  fixed,
  set thousands separator={\,},
  1000 sep in fractionals,
}

\newcolumntype{C}[1]{>{\centering\arraybackslash}m{#1}}
\newcolumntype{R}[1]{>{\raggedright\arraybackslash}m{#1}}
\newcolumntype{L}[1]{>{\raggedleft\arraybackslash}m{#1}}

\newcommand{\delete}[1]{\xspace}

\title{A 3D benchmark problem for crack propagation in brittle fracture}

\author[1]{L. Hug\thanks{lisa.hug@tum.de, Corresponding Author}}
\author[1]{S. Kollmannsberger}
\author[2]{Z. Yosibash}
\author[1,3]{E. Rank}

 \affil[1]{Chair for Computation in Engineering,
 Technische Universit\"at M\"unchen,
  Arcisstr. 21, 80333 M\"unchen, Germany}
 \affil[2]{School of Mechanical Engineering, Tel-Aviv University, 69978 Ramat-Aviv, Israel}
 \affil[3]{Institute for Advanced Study, Technische Universit\"at M\"unchen, Lichtenbergstr. 2a, 85748 Garching, Germany}

\newcommand{\journal}{Some Journal}
\newcommand{\publicationDate}{\today}

\usepackage{fancyhdr}
\date{}
\fancypagestyle{plain}{%
  \fancyhf{}%
  \fancyfoot[L]{
  \vspace{-1.25cm} 
  \footnotesize{Preprint submitted to \textit{\journal{}}  \hfill
  \publicationDate
  \\
  }} }

\crefname{figure}{Fig.}{Fig.}
\crefname{equation}{Eq.}{Eq.}
\crefname{table}{Tab.}{Tab.}

\drawboundingboxtrue
\drawboundingboxfalse 

\newcommand*{\figref}[2][]{%
	\hyperref[{fig:#2}]{%
		Fig.~\ref*{fig:#2}%
		\ifx\\#1\\%
		\else
		\,#1%
		\fi
	}%
}
\definecolor{changes}{RGB}{0,0,0}
\definecolor{changez}{RGB}{0,0,0}

\begin{document}  

\normalem
\maketitle  
  
\vspace{-1.5cm} 
\hrule 
\section*{Abstract}
\textcolor{changez}{W}e propose a full 3D benchmark problem for brittle fracture based on experiments as well as a validation in the context of phase-field models. The example consists of a series of four-point bending tests on graphite specimen\textcolor{changez}{s with} sharp V-notches at different inclination angles. This simple setup leads to a mixed mode \textcolor{changez}{(}I + II + III\textcolor{changez}{)} loading which results in complex yet stably reproducible crack surfaces. The proposed problem is \textcolor{changes}{well} suited for benchmarking numerical methods for brittle fracture and allows \textcolor{changez}{for a quantitative comparison of} failure loads and \textcolor{changez}{propagation paths as well as} initiation angles and the fracture surface. For evaluation of the crack surfaces image-based 3D models of the fractured specimen are provided along with experimental and numerical results. In addition, measured failure loads and \textcolor{changez}{computed} load-displacement curves are given. To demonstrate the applicability of the benchmark problem, we show that for a phase-field model based on the \textcolor{changes}{Finite Cell Method} and multi-level $hp$-refinement the \textcolor{changes}{complex} crack surface as well as the failure loads can be \textcolor{changez}{well} reproduced.
 \vspace{.2cm} 
\vspace{0.25cm}\\
\noindent \textit{Keywords:} Brittle fracture, benchmark, verification and validation, phase-field modeling, multi-level $hp$-adaptivity, Finite Cell Method 
\vspace{0.35cm}
\hrule 
\vspace{0.15cm}
\captionsetup[figure]{labelfont={bf},name={Fig.},labelsep=colon}
\captionsetup[table]{labelfont={bf},name={Tab.},labelsep=colon}
\tableofcontents
\vspace{0.5cm}
\hrule 

 \section{Introduction} \label{sec:intro}
\textcolor{changez}{N}umerical simulation of crack propagation is \textcolor{changez}{of major engineering importance} and a challenging problem. The main difficulty stems from the inherently discontinuous nature of the crack \textcolor{changez}{imposing} difficulties to numerical treatment. Consequently, a \textcolor{changes}{large number} of numerical approaches for fracture simulation have been proposed ranging from extended finite element methods (\cite{moes1999finite}, \cite{bely1999elastic}) and cohesive zone models \cite{ortiz1999finite} to diffusive approaches including phase-field models (\cite{fran1998revisiting}, \cite{bourd2000}) while research on new methods is continuing. \\
The phase-field approach to fracture overcomes the challenges faced by numerical modeling of crack propagation in a conventional finite element setting by replacing the discontinuous crack with a smoothed approximation. The crack is represented using a continuous scalar-variable, the so-called phase-field, which interpolates smoothly between fully broken and undamaged regions. This regularization of the sharp crack is based on a length-scale parameter which governs the extent of diffusion and determines the numerical width of the crack. For small length-scale parameters, a very fine mesh is needed in the vicinity of the crack to fully resolve the high gradients in the crack profile. For large domains and complex crack patterns in three dimensions this quickly becomes computationally expensive if a fixed mesh is used. Consequently, a range of numerical frameworks for phase-field brittle fracture have been developed which implement adaptive refinement to reduce \textcolor{changes}{the} computational effort. Within the isogeometric analysis framework different adaptive formulations based on hierarchical B-splines or NURBS have been proposed by e.g. \cite{borden2014higher}, \cite{hennig2016bezier} and \cite{hesch2016isogeometric}. An alternative approach based on multi-level hp-refinement using integrated Legendre polynomials was presented in \cite{nagaraja:18}. \\ \textcolor{changez}{T}o compare and evaluate different numerical \textcolor{changes}{approaches} \textcolor{changez}{for} simulat\textcolor{changez}{ing} complex 3D crack scenarios, benchmark problems are of great importance. 
In the brittle fracture community, popular benchmark examples include the single-edge notched tension (mode I) and shear test (mode II). These problems are particularly appealing due to their simple setup and thus often used for a first two-dimensional verification of newly implemented methods \textcolor{changes}{(\cite{bourd2000}, \cite{miehe2010phase}, \cite{ambati2015review})}. For a more complex mixed-mode I + II crack propagation \cite{galvez1998mixed} proposed a four-point bend of notched beams under the action of two independent actuators. \cite{rethore2018PMMA} presented a mixed mode crack propagation experiment on PMMA providing comprehensive data obtained by digital image correlation (DIC). Further 2D validations with experiments can be found in \cite{dally2017phase}, \cite{nguyen2015phase}. Complex, however in-plane 3D loading tests are presented in \cite{carpiuc2017complex} performed on specimen with a single notch under combined tensile shear and in-plane rotation inspired by the Nooru-Mohamed test (\cite{nooru1993mixed}, \cite{nooru1993experimental}). Three-point bending tests on a skew notched beam were investigated in \cite{citarella2008comparison} and recently in \cite{shao2019adaptive}. \cite{nguyen2016initiation} performed compression tests on concrete samples to compare 3D micro-cracking initiation and propagation in heterogeneous materials based on XR-$\mu$Ct images.\\
To \textcolor{changez}{the best of our} knowledge, no 3D benchmark problem for brittle fracture has been proposed which allows for a quantitative comparison of crack nucleation and propagation featuring a simple geometry and problem setup. To fill this gap, we propose a benchmark example based on a range of mixed mode I + II + III loading tests conducted in \cite{yosibash:16}, where rectangular bars containing a sharp V-notch at different inclination angles were loaded to fracture and fracture inclination angles as well as fracture loads were measured. We extend this data by 3D models of the fractured specimen in form of image-based point clouds, which allow\textcolor{changez}{s} for a 3D surface comparison of the whole crack surface. First numerical results for the proposed problem are presented which have been obtained using a phase-field model for \textcolor{changes}{simulation of} brittle fracture. The model extends the 2D implementation presented in \cite{nagaraja:18} to 3D and combines the Finite Cell Method with multi-level hp-refinement \textcolor{changez}{for} an efficient implementation which can easily be applied to complex geometries.\\ 
The paper is organized as follows: in Section \ref{sec:benchmark} setup and experimental results of the proposed benchmark problem are presented along with the definition of comparison quantities. Section \ref{sec:phasefield-theory} introduces the phase-field model for brittle fracture and its numerical implementation, followed by the numerical results in Section \ref{sec:results}. Finally, a summary is given in Section \ref{sec:conclusions}.  \textcolor{changes}{We provide geometric models of the fractured specimen together with the results of our simulations as supplementary data for download}.

 \section{The Benchmark: V-notched Specimen} \label{sec:benchmark}
The validation example presented in this section is based on experiments and results presented in \cite{yosibash:16} who developed a three-dimensional failure initiation criterion for brittle materials containing a V-notch. In a series of four-point bending experiments PMMA, Graphite and Macor V-notched specimen of different geometry \textcolor{changez}{were} fractured under mixed mode loading and failure loads as well as fracture initiation angles \textcolor{changez}{were} measured. For the proposed benchmark only the Graphite specimen\textcolor{changez}{s} are considered.
\begin{figure}[!b]
	\centering
	\includegraphics[width=0.6\textwidth]{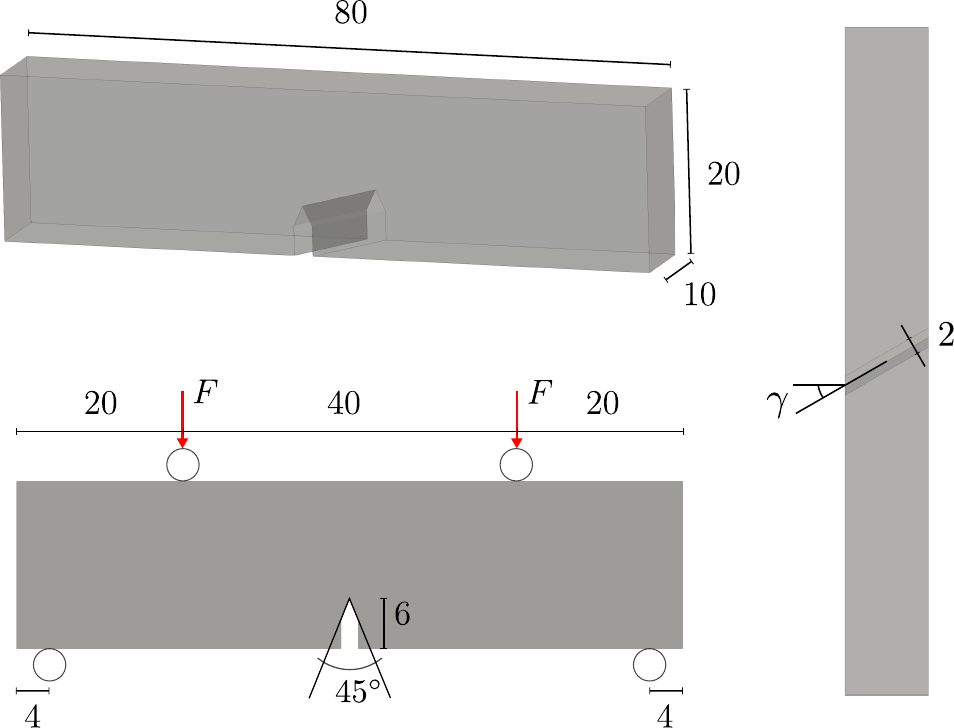}
	\caption{Geometry and Boundary conditions for V-notched specimen following \cite{yosibash:16}, \textcolor{changez}{(all dimensions in mm)}}
	\label{fig:exp-geometry}
\end{figure}
\subsection{Experimental Setup}
\begin{figure}[!t]
	\centering
	\includegraphics[width=0.55\textwidth]{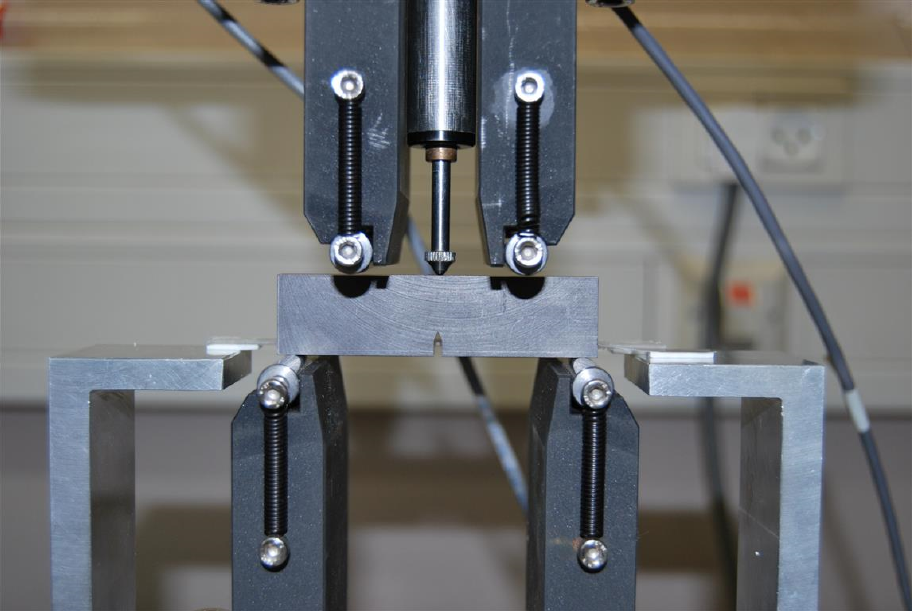}
	\caption{Experimental setup for the four-point bending experiments (\cite{yosibash:16}).}
	\label{fig:exp-setup}
\end{figure}

Four-point bending experiments were conducted on specimens with a\textcolor{changez}{n inserted} V-notch at three different inclination angles $\gamma = 0 \degree, 30 \degree $ and $45 \degree$. The geometry of the specimen and the experimental setup are shown in \cref{fig:exp-geometry} and \cref{fig:exp-setup}, respectively. The specimen are manufactured from graphite with dimensions of $80 \times 20 \times 10$ mm and a notch height of $h = 6$ mm. The notch has an opening angle of $45 \degree$ and was manufactured with a width of $2$ mm and a tip radius of $\rho = 50\,\mu \textrm{m}$. Placed on two cylinders the specimen were loaded under a constant displacement rate of  \mbox{$1$ mm$/$min}. Material properties have been determined as $E\,=\,12.44\,\, \textrm{kN}/\textrm{mm}^2$, \mbox{$G_c\, =\, 1.18 \times 10^{-4}$\, kN$/$mm}, $\nu\, =\, 0.2$ and $\sigma_c\, =\, 48$ MPa as described in \cite{yosibash:16}.
\subsection{Comparison Quantities}
In order to assess quality of the simulation results the critical load to failure, fracture initiation angles and the resulting fracture surface are \textcolor{changez}{compared to the experimental observations}.
\subsubsection{Failure Loads}
The experimental failure loads are summarized in Tab. \ref{tab:fractureloads}. A total number of eleven graphite specimen\textcolor{changez}{s have} been fractured including one specimen with $\gamma=0\degree$, five with $\gamma=30\degree$ and five with $\gamma=45\degree$. The average value of all experiments for one geometry will be used as \textcolor{changez}{the} reference value. The relative deviation in \textcolor{changez}{percentage} is given as the average deviation divided by the mean\textcolor{changez}{:} $2.64\, \%$ for the $\gamma=30\degree$ geometry and $1.03\, \%$ for the $\gamma=45\degree$ case. As pointed out in \cite{yosibash:16}, due to the difference in loading modes for the three geometries for higher inclination angles $\gamma$ the load to fracture increases which is clearly visible in the experiments. The numerical results for the failure load will be obtained from a load-displacement curve in a quasi-static simulation as the force associated to the first local maximum, see Section \ref{sec:res-failureloads}. 
\begin{table}[b]
	\centering
	\setlength{\tabcolsep}{5pt}
	\renewcommand{\arraystretch}{1.3}
	\begin{tabular}{llll} \toprule
		$\gamma$ [$\degree$] & avg. force [kN] & mean deviation [$\%$] & \#spec.\\ \midrule
		$0 $ & \quad $0.668 $ & - & 1\\
		$30 $ & \quad$0.846$& $2.64\,$ & 5\\
		$45 $ & \quad$1.028$& $1.03\,$ & 5\\ \bottomrule
	\end{tabular}
	\caption{Experimental results for the failure loads from \cite{yosibash:16}.}
	\label{tab:fractureloads}
\end{table}
\subsubsection{\textcolor{changez}{Fracture} Initiation Angles}
\begin{figure}[!b]
	\centering
	\includegraphics[width=0.95\textwidth]{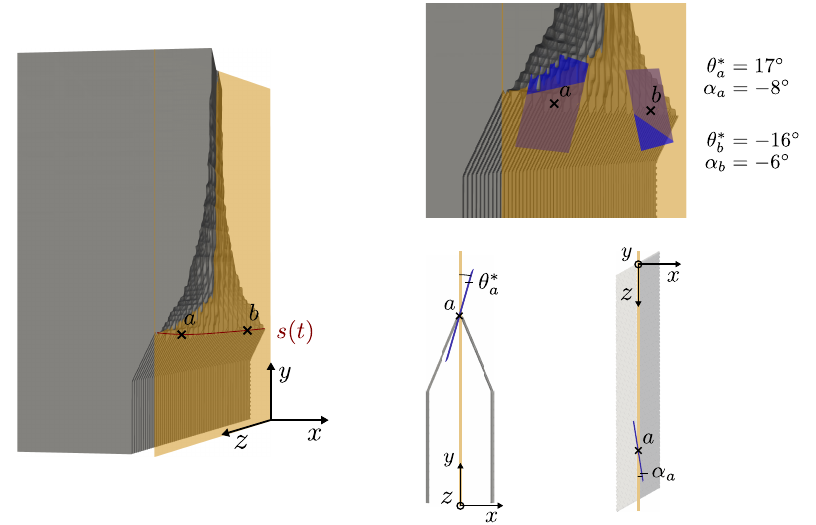}
	\caption{\textcolor{changes}{\textcolor{changez}{Half broken specimen} with definition of the initiation angles $\alpha$ and $\theta^{*}$. Rotated coordinate system and crack initiation curve $s(t)$ with two evaluation points $a$ and $b$, left. Tangential planes in $a$ and $b$ (blue) and definition of $\alpha$ and $\theta^{*}$ with respect to the V-notch bisector plane (orange), right.}}
	\label{fig:initiationAngDef}
\end{figure}
\textcolor{changes}{Fracture initiation is compared by means of two rotation angles, $\alpha$ and $\theta^{*}$, which describe the orientation of the fracture surface along the notch. Both angles are defined relative to the V-notch bisector plane with respect to a rotated coordinate system, which is chosen for each geometry such that the $z$-axis is aligned with the notch \textcolor{changez}{tip edge}. The rotation angles are evaluated for different $z$-coordinates along the crack initiation curve $s(t)$ as visualized in \cref{fig:initiationAngDef}, left. For each point on $s(t)$ a tangential plane to the fracture surface can be defined. $\alpha$ and $\theta^{*}$ are then obtained as the angles which describe the rotation of the tangential plane with respect to the V-notch bisector plane, see \cref{fig:initiationAngDef}, right.} 
The first angle, $\theta^{*}$, describes a counter-clockwise rotation around the $z$-axis, while the second angle, $\alpha$, describes a counter-clockwise rotation of the fracture surface around the $y$-axis. For a crack initiating straight along the notch tip parallel to the $y$-axis it holds $\alpha\,=\,0$ and $\theta^{*}=0$. Experimental \textcolor{changez}{observations} are presented along the numerical results in Figure \ref{fig:initiationAngles} and discussed in detail in Section \ref{sec:res-initangles}.
\subsubsection{Fracture Surface}
\textcolor{changez}{T}o assess the quality of the \textcolor{changez}{computed} crack surface\textcolor{changes}{,} image-based 3D representations of the fractured specimens are created. Based on a sequence of images recorded along a circular path a modified bundle adjustment algorithm (\cite{triggs1999bundle}, \cite{kudela2018image}) is used to generate a 3D point cloud of the specimen. Using a suitable metric the distances between the obtained reference point set and the 3D surface model from the simulation can be evaluated and compared. For two non-empty set\textcolor{changes}{s} of points $\mathcal{A}$ and $\mathcal{B}$, we define the distance between a point $a\in\mathcal{A}$ and the set $\mathcal{B}$ as 
\begin{equation}
\label{eq.dist-measure}
d(a,\mathcal{B}) = \underset{b\in \mathcal{B}}{\textrm{min}} \lVert a-b \rVert\,,
\end{equation}
where $\lVert \cdot \rVert$ denotes the $\textcolor{changez}{L}^2$-norm. A commonly used distance measure is the Hausdorff distance which is defined as 
\begin{equation}
\label{eq:H-hd}
H(A,\mathcal{B}) = {\textrm{max}}\left(h(\mathcal{A},\mathcal{B}), h(\mathcal{B},\mathcal{A})\right)\,,
\end{equation}
where
\begin{equation}
\label{eq:hd}
h(\mathcal{A},\mathcal{B}) = \underset{a\in \mathcal{A}}{\textrm{max}}\,d(a,\mathcal{B})\,,
\end{equation}
is the directed Hausdorff distance. The modified Hausdorff distance (MHD) as proposed by \cite{dubuisson1994modified} promises superior properties for object matching and is obtained as the arithmetic mean of all nearest neighbor distances following
\begin{equation}
\label{eq:H-mhd}
H_{\textrm{MHD}}(\mathcal{A},\mathcal{B}) = {\textrm{max}}\left(h_{\textrm{MHD}}(\mathcal{A},\mathcal{B}), h_{\textrm{MHD}}(\mathcal{B},\mathcal{A})\right)\,,
\end{equation}
where
\begin{equation}
\label{eq:mhd}
h_{\textrm{MHD}}(\mathcal{A},\mathcal{B}) = \frac{1}{N_a}\sum_{a \in \mathcal{A}} d(a,\mathcal{B})\,
\end{equation}
and $N_a$ is the number of points in set $\mathcal{A}$. 
For evaluating the distances $d(a,\mathcal{B})$ the software \textcolor{changes}{\cite{cloudcomp}} is used. 
\subsection{Provided Data \textcolor{changez}{for Benchmarking}}
Tables containing failure loads and fracture initiation angles for all specimen can be found in \cite{yosibash:16}. Point clouds of specimen no.$\,$6 ($\gamma=30\degree$) and specimen no.$\,$8 ($\gamma=45\degree$) are available at Mendeley Data (http://dx.doi.org/xyz) as '.ply' files. In addition, the crack surfaces obtained with the phase-field model presented in Section \ref{sec:results} are provided for all three inclination angles as '.stl' files as well as the numerical load-displacement data. 
\begin{figure}[b]
	\centering
	\includegraphics[width=0.84\textwidth]{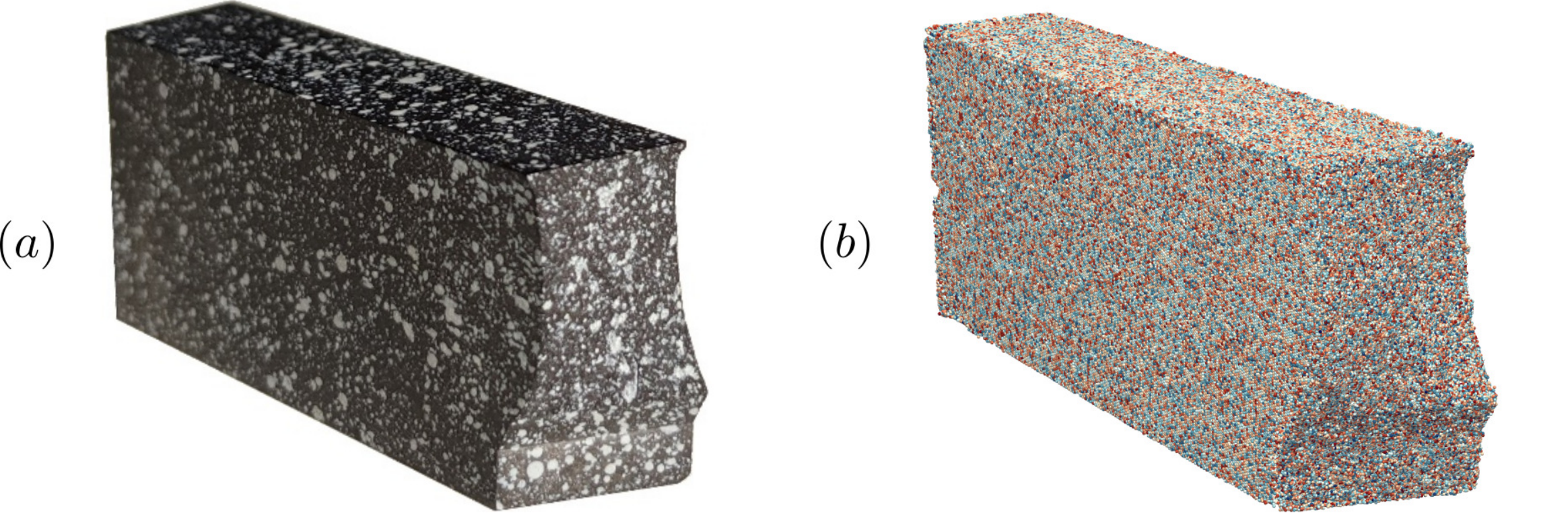}
	\caption{Photo of the fractured specimen $(a)$ and constructed point cloud $(b)$ for the $30 \degree$ geometry. The specimen was sprinkled with white color to enhance surface perception for the feature extraction algorithm.}
	\label{fig:specpointcloud}
\end{figure}

 \section{Phase-Field Formulation} \label{sec:phasefield-theory}
The basic theories \textcolor{changez}{of brittle fracture} were first introduced by \cite{griffith1921} and later \cite{Irwi1958elasticity}, who formulated a criterion for crack propagation based on an energy balance: A crack propagates as soon as the elastic strain energy which is released in the process of crack extension exceeds a critical energy, the so-called fracture toughness $G_c$. Following Griffith's theory, $G_c$ describes the resistance of a material to crack formation and is equal to the surface energy required to form new crack surfaces. \textcolor{changes}{These theories only provide a criterion for crack propagation. A} variational approach based on energy minimization has been introduced by \cite{fran1998revisiting} and later regularized by \cite{bourd2000} to enable efficient numerical treatment. 
\subsection{Energy Functional and Variational Problem} 
\begin{figure}[b]
 	\centering
	\includegraphics[width=0.92\textwidth]{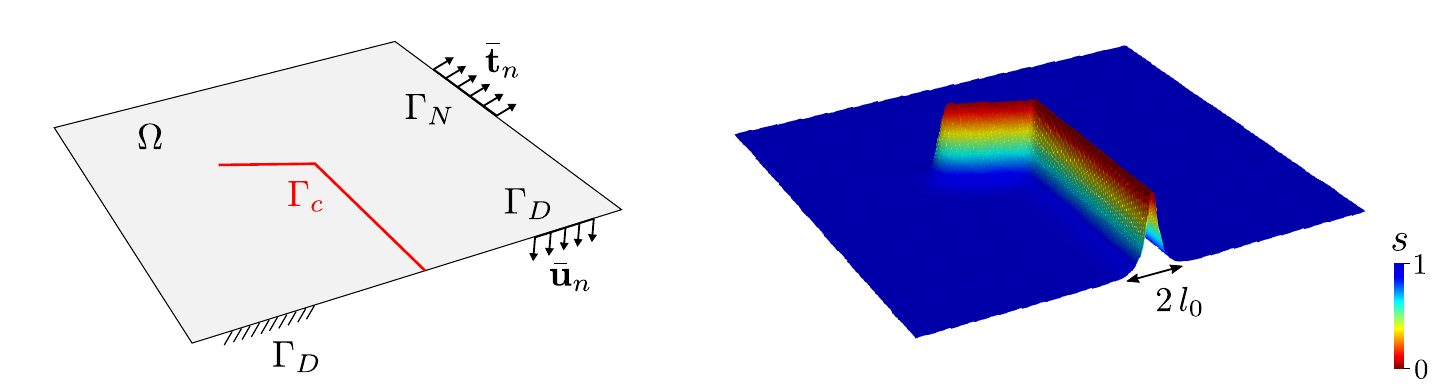}
	\caption{Sharp crack topology \textcolor{changes}{(left)} and phase-field regularized crack surface \textcolor{changes}{(right)} after \cite{gerasimov2018penalization}.}
	\label{fig:phasefield_setup}
\end{figure}

In the following, let $\Omega \subset \mathcal{R}^d$, $d=2,\,3$ be an open bounded domain with boundary $\partial \Omega$, which is cut by a set of discrete cracks $\Gamma_c$ as shown in \cref{fig:phasefield_setup}\textcolor{changes}{, left}. Furthermore, $\Gamma_D$ and $\Gamma_N$ refer to the non-overlapping parts of $\partial \Omega$ on which Dirichlet and Neumann boundary conditions are prescribed. A point in $\Omega$ is denoted by ${\boldsymbol x}$ and ${\boldsymbol u}({\boldsymbol x}), {\boldsymbol{\varepsilon}}({\boldsymbol x})$ and ${ \boldsymbol{\sigma}}({\boldsymbol x}) \in  \mathcal{R}^d$ are the displacement, strain and stress fields, respectively. Small deformations are assumed with the strain tensor given as $\boldsymbol{\varepsilon} = \frac{1}{2}\,(\nabla \boldsymbol{u} + \nabla^{\textbf{T}} \boldsymbol{u})$ and the elastic strain density as $\Psi({\boldsymbol{\varepsilon}})=\frac{1}{2}\,\lambda\,\textrm{tr}^2(\boldsymbol{\varepsilon}) + \mu\,\textrm{tr}(\boldsymbol{\varepsilon}^2)$, \textcolor{changes}{where $\lambda$ and $\mu$ are the Lam\'e constants}.\\
\cite{fran1998revisiting} introduced the variational formulation of brittle fracture as a minimization problem of the energy functional
\begin{equation}
\label{eq:fracfort_functional}
E({\textbf u}, \Gamma_{\textcolor{changes}{c}})\,=\, \int_{\Omega}\,\Psi(\boldsymbol{\varepsilon}) \textrm{d}\textcolor{changes}{x}\,+\,\int_{\Gamma_{\textcolor{changes}{c}}} G_{c} \textrm{d}\textcolor{changes}{s}\,.
\end{equation}
Here, the total potential energy consists of the energy stored in the solid and the surface energy dissipated in the fracture process creating the crack set $\Gamma_{\textcolor{changes}{c}}$. 
To facilitate numerical solution of (\ref{eq:fracfort_functional}), a regularized formulation (\cite{bourd2000}, \cite{bour2008}) was proposed which approximates the surface integral as a volume term yielding
\begin{align}
\label{eq:bourdin_reg_functional}
E_{l_0}({\textbf u}, s)\,=\, \int_{\Omega}\,g(s)\,\Psi(\boldsymbol{\varepsilon}) \textrm{d}{\textbf x}\,+\,\frac{G_{c}}{c_w}\int_{\Omega} \left( \frac{1}{2\,l_0} w(s) \,+\, 2\,l_0 |\nabla s|^2\right) \textrm{d}{\textbf x}\,
\end{align}
Here, a regularization length $l_0$ and a continuous scalar-variable $s\,\in\,[0,1]$, the phase-field parameter, are introduced. The sharp crack is smoothed and modeled as a continuous field which attains a value of one in undamaged regions and is zero along the crack (cf. \cref{fig:phasefield_setup}). The width of the transition zone is governed by the length-scale parameter $l_0$. In the limit $l_0 \rightarrow 0$ the regularized functional (\ref{eq:bourdin_reg_functional}) converges to the original formulation (\ref{eq:fracfort_functional}) in the sense of $\Gamma$-convergence, i.e. the global minimizers of $E_{l_0}({\textbf u}, s)$ converge to those of $E({\textbf u}, {\textcolor{changes}{\Gamma_c}})$ (\cite{alberti2000variational}). The degradation function $g(s)$ models the loss of stiffness in the damaged material and $c_w$ is a scaling parameter related to the energy dissipation function $w(s)$. 
As formulation (\ref{eq:bourdin_reg_functional}) may result in non-physical crack patterns and interpenetration of crack surfaces in compression, additive splitting of the elastic strain energy density $\Psi(\boldsymbol \varepsilon) = \Psi^+(\boldsymbol \varepsilon) + \Psi^-(\boldsymbol \varepsilon)$ was proposed by several authors, including \cite{amor2009regularized} and \cite{miehe2010thermodynamically}. Degradation acts only on the positive part of the energy density resulting in the slightly altered energy functional
\begin{align}
\label{eq:functional_w_split}
E_{l_0}({\textbf u}, s)\,=\, \int_{\Omega}\,g(s)\,\Psi^+(\boldsymbol \varepsilon) + \Psi^-(\boldsymbol\varepsilon)\textrm{d}{\textbf x}\,+\,\frac{G_{c}}{c_w}\int_{\Omega} \left( \frac{1}{2\,l_0} w(s) \,+\, 2\,l_0 |\nabla s|^2\right) \textrm{d}{\textbf x}\,.
\end{align}
In this contribution, the spectral split by \cite{miehe2010thermodynamically} is used.
\subsection{Governing Equations}
Following the variational principle minimizers of the functional \cref{eq:functional_w_split} can be found by solving the associated set of Euler-Lagrange equations. The strong form of the coupled system is obtained as 
\begin{subequations}
\label{eg:governeq}
\begin{align}
\textrm{div}(\bm{\sigma})\,+\,\rho\,{\bm b}&=0,\qquad  \textrm{where } \bm{\sigma} = g(s)\,\dfrac{\partial \Psi^+(\boldsymbol  \varepsilon)}{\partial \boldsymbol \varepsilon} + \frac{\partial \Psi^-( \boldsymbol \varepsilon)}{\partial \boldsymbol\varepsilon} \\[0.1cm]
-4\,l_0^2\,\Delta s\,+\frac{1}{2}\,w'(s)&=\,-\frac{c_w}{2}\,\frac{l_0}{G_c}\,g'(s)\,\mathcal{H}
\end{align}
\end{subequations}
subject to the boundary conditions
\begin{align}
\label{eq:bc}
{\textbf u} &= \bar{\textbf{u}}_n\quad &&\textrm{on }\Gamma_{D}, \\
{\boldsymbol{\sigma}}\cdot {\textbf{n}} &= \bar{\textbf{t}}_n &&\textrm{on }\Gamma_N, \\
 \nabla d \cdot {\textbf{n}} &= 0 &&\textrm{on }\Gamma_{D} \cup \Gamma_N. 
\end{align}
Following the approach proposed by \cite{miehe2010thermodynamically} irreversibility of the phase-field is ensured by introducing a history variable $\mathcal{H}$ defined as
\begin{align}
	\mathcal{H}(\textbf{x},t) \coloneqq \underset{t \in [0,T]}{\textrm{max}}\Psi^+(\boldsymbol \varepsilon(\textbf{x},t))\,,
\end{align} 
which replaces the positive part of the elastic strain density in the phase-field equation. That way, the equations are decoupled and the system of equations can be solved alternately applying a staggered solution scheme. Different choices for the degradation and energy dissipation functions are possible. Commonly used are the quadratic functions
\begin{align}
\begin{split}
\label{eq:degradation}
&g(s) = (1-\eta)s^2 + \eta\,,\\
&w(s) = 1- s^2\,,\, c_w = 2\,
\end{split}
\end{align}
where $\eta \approx 0$ is a small parameter ensuring numerical stability in case of a fully degraded material. An alternative choice is a cubic degradation function (\cite{borden2012isogeometric}) given as
\begin{align}
\label{eq:cubic_degrad}
g(s) = \phi\,(s^3 - s^2) + 3\,s^2- 2\,s^3\,,
\end{align}
where $\phi$ determines the slope of the degradation function for $s=1.0$ \textcolor{changes}{and is chosen as $10^{-4}$ for all computations}. Higher order degradation functions have the desirable property to lead to an approximately linear elastic behavior prior to fracture at the cost of solving a non-linear equation for the phase-field. 

\subsubsection{The Hybrid Model} \label{sec:hybridmodel}
The hybrid formulation was introduced by \cite{ambati2015review} to reduce the computational demand stemming from the non-linearity of the elastic equation in the anisotropic model. In comparison to the fully anisotropic model, the split is accounted for only in the phase-field equation. The elastic equation takes the classic isotropic form with the stress tensor being defined as
\begin{equation}
\label{eq:hybrid}
\bm{\sigma} = g(s)\,\dfrac{\partial \Psi(\boldsymbol  \varepsilon)}{\partial \boldsymbol \varepsilon}\,,
\end{equation}
which results in a linear elastic equation. Thus, the hybrid model promises comparable results to the anisotropic model while being computationally much cheaper (\cite{ambati2015review}). \cite{jeong2018phase} pointed out, that under combined tensile and shear loading the anisotropic formulation suffers from physically unrealistic behavior including further crack growth after complete failure and an overestimation of failure loads. In these cases, the hybrid formulation leads to improved results.
\subsubsection{Dynamic Fracture}\label{sec:dynamicmodel}
In \cite{borden2012phase}, the quasi-static model introduced above was extended to the dynamic case. The energy functional additionally accounts for the kinetic energy of the body and yields the following coupled system of equations
\begin{subequations}
   \label{eq:hybrid-dynamic}
	\begin{align}
		\label{eq:el-hybrid-dynamic}
\textrm{div}(\bm{\sigma})\,+\,\rho\,{\bm b}&=\rho\, {\bm \ddot{u}},\qquad  \textrm{where } \bm{\sigma} = g(s)\,\dfrac{\partial \Psi(\boldsymbol  \varepsilon)}{\partial \boldsymbol \varepsilon} \\[0.1cm]
-4\,l_0^2\,\Delta s\,+\frac{1}{2}\,w'(s)&=\,-\frac{c_w}{2}\,\frac{l_0}{G_c}\,g'(s)\,\mathcal{H}\,.
	\end{align}
\end{subequations}
Here $\rho$ denotes the density of the material and the hybrid formulation introduced in Section \ref{sec:hybridmodel} is used.
\begin{figure}[b]
	\centering
	\includegraphics[width=0.75\textwidth]{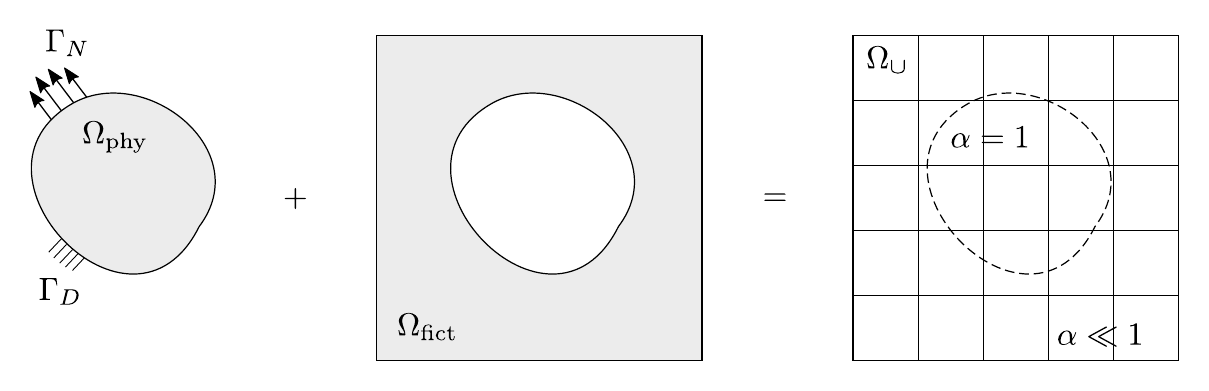}
	\caption{Embedding concept of the Finite Cell Method following \cite{parvizian2007finite}.}
	\label{fig:fcm}
\end{figure}
\subsection{Numerical Description}
The numerical framework extends the 2D implementation presented in \cite{nagaraja:18} to 3D and combines a phase-field model with the FCM (\cite{parvizian2007finite}) and multi-level $hp$-adaptive refinement (\cite{zander2015multi}). The adaptive refinement technique makes it possible to efficiently resolve the locally occurring high gradients in the phase-field even in the case of small length-scale parameters. Implementation in terms of the FCM allows to simulate complex geometries without the need to generate a boundary conforming mesh.
\subsubsection{The Finite Cell Method}
As an immersed boundary method, the basic idea of FCM consists in embedding the physical domain $\Omega_{phy}$ into a geometrically larger domain of simple shape, \textcolor{changes}{$\Omega_{\cup} = \Omega_{phy} \cup \Omega_{fict}$}, see \cref{fig:fcm}. This domain can easily be meshed using a structured grid which shifts the task of representing the actual geometry from discretization to the integration. In order to differentiate \textcolor{changes}{between $\Omega_{phy}$ and the added fictitious part $\Omega_{fict}$} an indicator function $\alpha({\boldsymbol x})$ is introduced:
\begin{equation}
\alpha({\boldsymbol x})\,=\,\begin{cases} 1.0\,, \quad \forall {\boldsymbol x} \in \Omega_{phy}, \\ \alpha_{FCM}\,, \quad \forall {\boldsymbol x} \in \textcolor{changes}{\Omega_{fict}}, \end{cases}
\end{equation} 
where $\alpha_{FCM}$ is a small parameter greater than but unequal to zero to avoid ill-conditioning of the stiffness matrix and ensure numerical stability (\cite{parvizian2007finite}). By multiplying the weak form with $\alpha({\boldsymbol x})$ contributions of the fictitious domain are penalized imitating the behavior of a very soft material. A crucial point is accurate numerical integration of the cells cut by the boundary of the original domain which feature a discontinuity. As Gauss quadrature shows poor convergence for non-smooth functions a number of alternative integration schemes has been proposed in the context of FCM, see \cite{hubrich2017numerical}. In this contribution the octree subdivision approach is used which was first combined with FCM in \cite{duster2008finite}. Here, integration accuracy is improved by a successive subdivision of cut cells up to a fixed partitioning depth.
\subsubsection{Multi-level $hp$-FEM}
\begin{figure}[!t]
	\centering
	\includegraphics[width=0.99\textwidth]{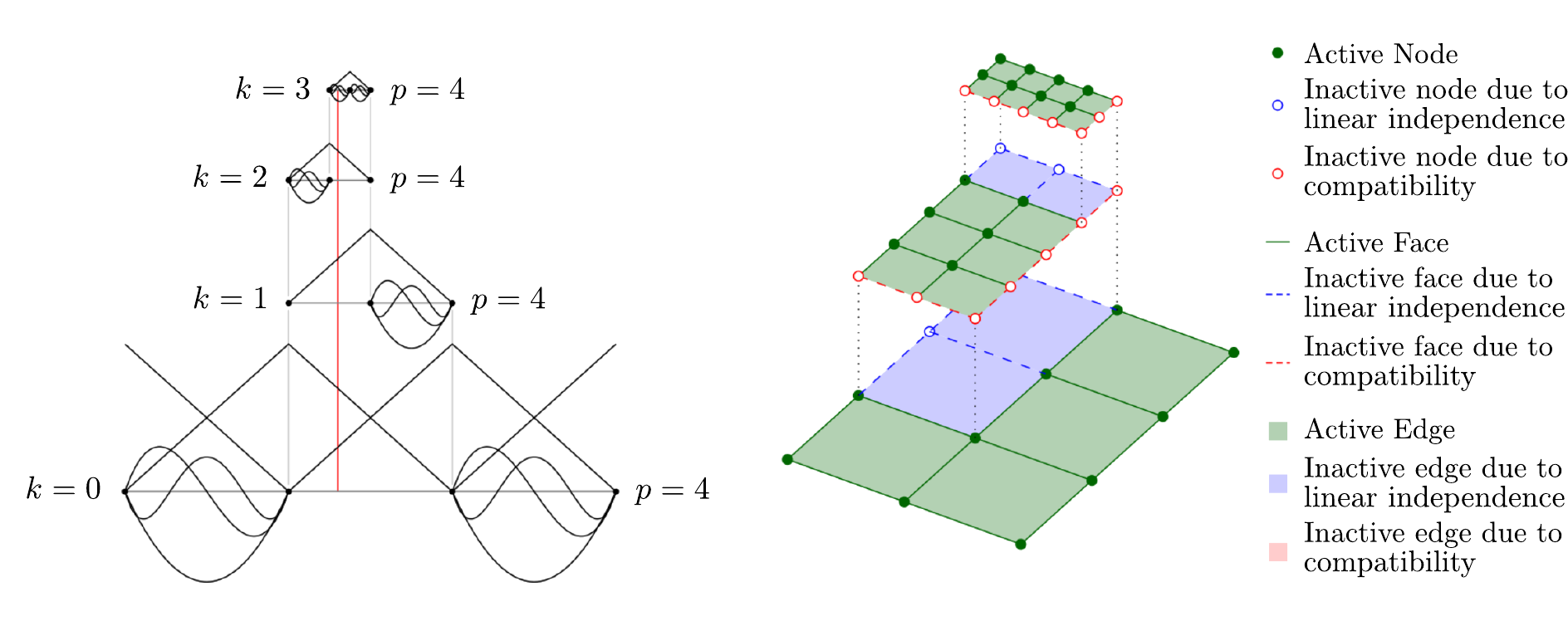}
	\caption{Conceptual idea of the multi-level $hp$-FEM for 1D (left) and 2D (right) following \cite{zander2015multi}.}
	\label{fig:multi-hp-fem}
\end{figure}
First introduced in \cite{zander2015multi}, the basic idea of this refinement technique is to replace the refine-by-replacement paradigm with a superposition approach. As visualized for a 1D and 2D example in \cref{fig:multi-hp-fem}, the base mesh is superposed with high-order overlay meshes in regions of interest, where $k$ is the refinement depth and $p$ refers to the ansatz order. This allows to avoid hanging nodes and the related implementational difficulties while the same approximation quality as in conventional $hp$-FEM is achieved (\cite{di2016easy}). To this end, two requirements need to be fulfilled: the compatibility and linear independence of the shape functions. Homogeneous Dirichlet boundary conditions are imposed on each layer of the superposed overlay mesh to maintain global $C^0$-continuity and thus compatibility of the basis functions. This is achieved by deactivating all nodal, edge and face modes on the boundaries of the overlay meshes. In addition, to guarantee linear independence of basis functions all topological components with active sub-components are deactivated. Both procedures are illustrated in \cref{fig:multi-hp-fem}. The multi-level $hp$-refinement technique allows for a flexible and dynamic update of the discretization throughout the simulation.
\subsubsection{Weak Form and Solution of the Coupled Problem}
Choosing test functions $\boldsymbol w(\boldsymbol x) \in H^1_0(\Omega)$ and $ q \in H^1(\Omega)$ for the elastic and phase-field problem, respectively, the weak form of the coupled quasi-static problem (Eq. \ref{eq:hybrid} ) for FCM is derived as
\begin{subequations}
\begin{equation}
\begin{aligned}
\left(\boldsymbol{\sigma}, \nabla \boldsymbol{w}\right)_{\Omega_{phy}} + (\alpha_{FCM} \,\boldsymbol{\sigma}, \nabla \boldsymbol{w})_{\Omega_{fict}} + (\beta \, \boldsymbol{u}, \boldsymbol{w})_{\Gamma_{D}} = (\rho\,\boldsymbol{b}, \boldsymbol{w})_{\Omega_{phy}} + (\boldsymbol{h},\boldsymbol{w})_{\Gamma_{D}} + (\beta \, \boldsymbol{g}, \boldsymbol{w})_{\Gamma_{D}}\,,
\end{aligned}
\end{equation}
\vspace{-1em}
\begin{equation}
\begin{aligned}
\left(\left[ \frac{4\,l}{G_c}(1-\eta)\mathcal{H}+1\right]\,s,q\right)_{\Omega_{phy}} + \left(\alpha_{FCM} \left[ \frac{4\,l}{G_c}(1-\eta)\mathcal{H}+1\right]\,s,q\right)_{\Omega_{fict}}  + \left(4\,l^2\,\nabla s, \nabla q \right)_{\Omega_{phy}}\\[2mm]  + \left(\alpha_{FCM}\, 4\,l^2\,\nabla s, \nabla q \right)_{\Omega_{fict}} = (1,q)_{\Omega_{phy}}\,,
\end{aligned}
\end{equation}
\end{subequations}
\textcolor{changes}{where $(\cdot\,, \cdot)$ denotes the $L^2$ scalar product}. Dirichlet boundary conditions for the elastic problem are imposed weakly using the penalty method with $\beta$ being the penalty parameter. For the discretization integrated Legendre polynomials are used as basis functions for the FE test and trial spaces \textcolor{changes}{(see \cite{babuska1981p})}, alternative formulations for FCM include B-splines (\cite{schillinger2012small}) and implementations in the framework of isogeometric analysis. The discretized problem is solved using a staggered solution approach. To \textcolor{changes}{control} the number of staggered iterations in each displacement step an energy-based criterion similar to the one proposed in \cite{gerasimov2016line} is implemented. The iterative procedure is stopped if for displacement step $i$ the number
\begin{equation}
S_{\textrm{tol}, i} = \frac{E_{i-1} - E_i}{E_{\textrm{init}} - E_i}\,
\end{equation}
falls below a certain threshold value $\varepsilon_{\textrm{stag}}$, or alternatively, if a set maximum number of staggered iterations $n_{\textrm{max, stag}}$ is reached. \textcolor{changez}{Here, $E_i = \sqrt{\frac{1}{n_{dof}}\sum_{j=1}^{n_{dof}} \frac{1}{2}\,\boldsymbol{u}_i\,\boldsymbol{K}_i\,\boldsymbol{u}_i}\,$.} 
 \section{Numerical Results} \label{sec:results}
In this section, the proposed benchmark problem is \textcolor{changes}{analyzed} using a phase-field model for brittle fracture based on the FCM and multi-level $hp$-refinement. Two different hybrid models will be used: the quasi-static model presented in Section \ref{sec:hybridmodel} and the dynamic model introduced in Section \textcolor{changes}{\ref{sec:dynamicmodel}}. \textcolor{changes}{The} quasi-static model is used to obtain the failure loads in Section \ref{sec:res-failureloads}, while a dynamic simulation is used to extract the fracture surface analyzed in Sections \ref{sec:res-surface} and \ref{sec:res-initangles}. The quasi-static model allows for a straight forward extraction of the load-displacement curves, while only the dynamic model is able to produce a fully cracked specimen under the present loading conditions.
\begin{figure}[b!]
	\centering
	\includegraphics[width=0.97\textwidth]{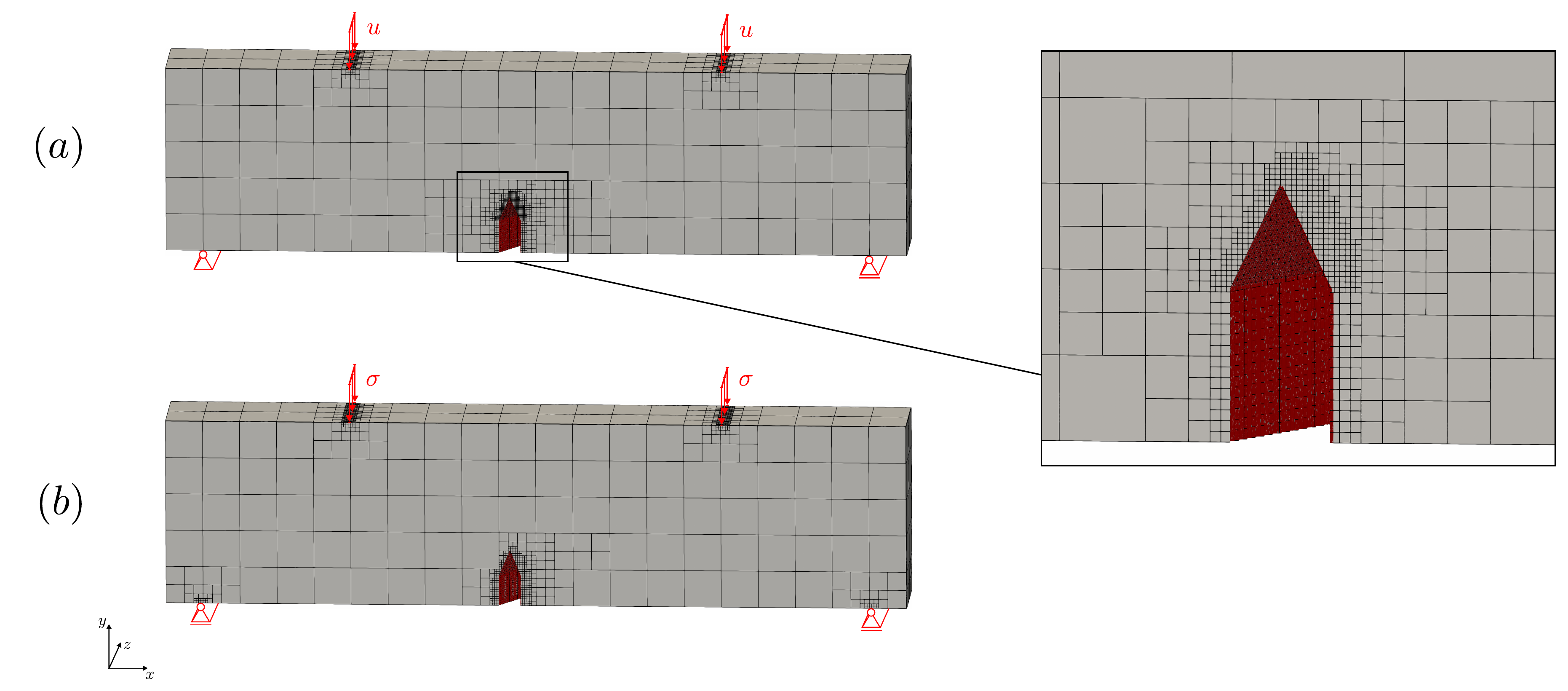}
	\caption{Domain with initial refinement and boundary conditions for the quasi-static (a) and the dynamic model (b). }
	\label{fig:simsetup}
\end{figure}
\subsection{Problem Setup} \label{sec:prob-setup}
Setups for the quasi-static and the dynamic simulation are shown in Fig. \ref{fig:simsetup}. In both cases, the domain sized $80 \textrm{ mm}\,\times\,20 \textrm{ mm}\,\times 10 \textrm{ mm}$ is initially discretized with $20\,\times\,5\,\times 2$ elements and refined towards the notch and the boundary conditions with a refinement depth of $k\,=\,4$. Note here, that the notch geometry is not resolved explicitly with the grid but represented using the FCM. In order to properly resolve the phase-field length scale parameter $l_0$ in the case of crack nucleation, in the quasi-static setup the upper part of the notch is additionally refined such that the element size equals $l_0$. For the dynamic case a resolution of $h=2\,l_0$ proves sufficient, as crack propagation is independent of $l_0$ if the latter is chosen small enough and geometrically meaningful. With this resolution the tip radius is neglected and a sharp notch is assumed. The mesh is refined dynamically in each load step based on the local phase-field solution using a threshold value of $0.7$, i.e. an element is refined if $s \leq 0.7$. Material and numerical parameters for both setups are listed in \cref{tab:simparams}. The material parameters are adopted from \cite{yosibash:16}. In order to prevent nucleation of cracks where the boundary conditions are applied, $G_c$ is increased by a factor of $10^{6}$ in these regions. The tolerance parameter $\varepsilon_{\textrm{stag}}$ was chosen small enough such that influence on the results was negligible. 

\begin{table}[t]
	\centering
	\small
	\setlength{\tabcolsep}{3.2pt}
	\renewcommand{\arraystretch}{1.6}
	\begin{tabular}{c|c|c|c|c|c|c|c|c|c}
		\toprule[0.25mm]
		model & $E\,[\frac{\textrm{kN}}{\textrm{mm}^2}]$ & $G_c\,[\frac{\textrm{kN}}{\textrm{mm}}]$ & $\sigma_c\,[\textrm{MPa}]$ & $\nu\,[-]$
		& $\rho\,[\frac{\textrm{kg}}{\textrm{m}}]$ & $l_0\,[{\textrm{m}}]$ & $\alpha_{FCM}\,[-]$ & $\varepsilon_{\textrm{stag}}\,[-]$ 
		& $n_{\textrm{stag}}\,[-]$ \\ \midrule
		$(a)$ & \multirow{ 2}{*}{$12.44 $} &  \multirow{ 2}{*}{$\begin{cases} 10^{-3}\quad \textrm{if}\,\, 35.0 \leq x \leq 55.0\\ 1000\quad\textrm{else} \end{cases}$} \hspace{-1em} & \multirow{ 2}{*}{$48.0 $} & \multirow{ 2}{*}{$0.2$} & $-$ & \multirow{ 2}{*}{$0.125$} & \multirow{ 2}{*}{$10^{-4}$} & \multirow{ 2}{*}{$5^{-3}$} & 35\\
		$(b)$ & & & & &1.066 & & & &18\\ \bottomrule[0.25mm]	
	\end{tabular} \\[3.2mm]
	\caption{Parameters for the quasi-static (a) and dynamic model (b).}
	\label{tab:simparams}
\end{table}

\subsection{Failure Loads}\label{sec:res-failureloads}
\begin{figure}[!b]
	\centering
	\includegraphics[width=0.85\textwidth]{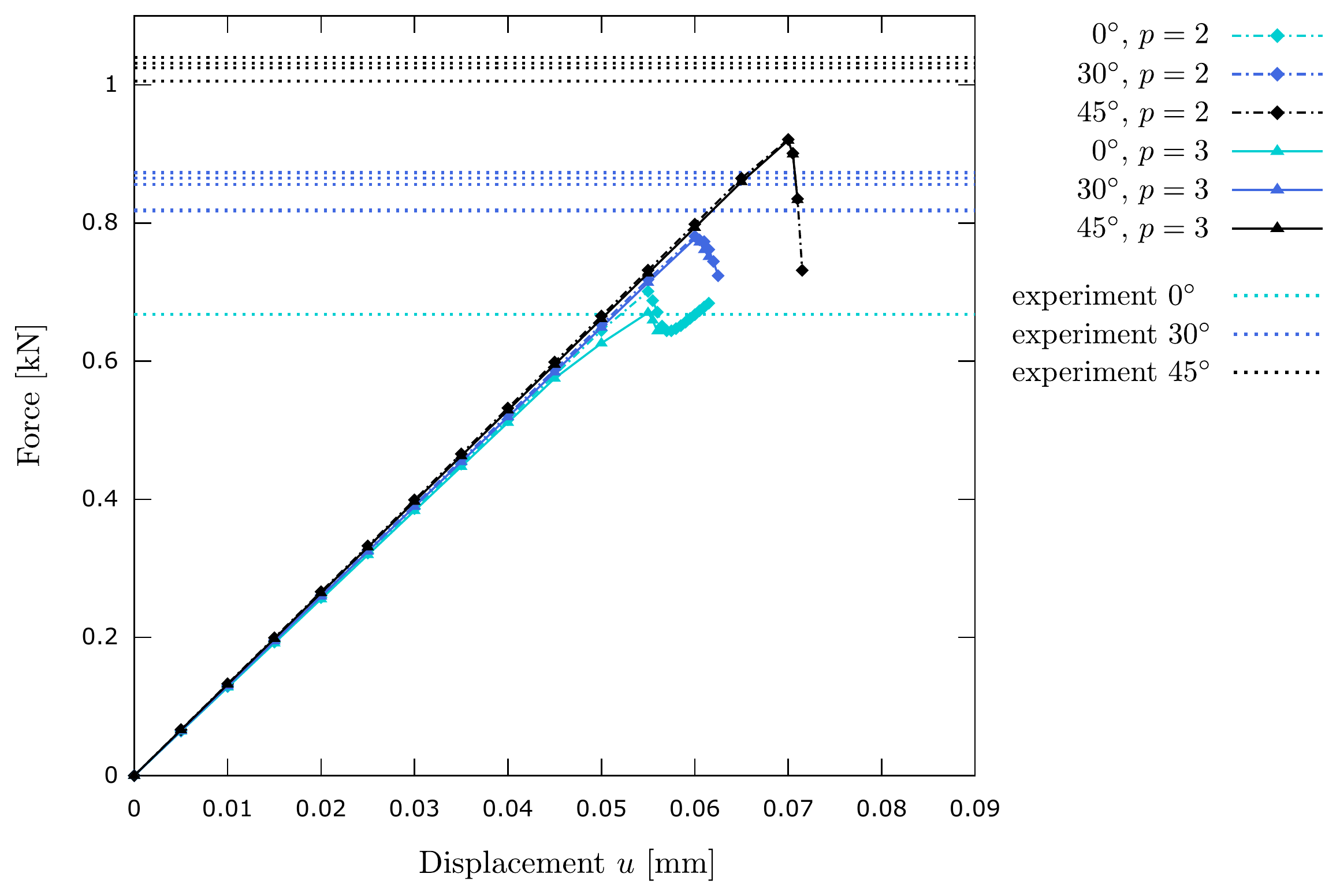}
	\caption{Load-displacement curve for graphite specimen with different inclination angles and experimental failure loads for ansatz oders $p=2$ and $p=3$.}
	\label{fig:failureloads}
\end{figure}
Quasi-static simulations are used to generate load-displacement curves for the three geometries. A cubic degradation function (Eq. \ref{eq:cubic_degrad}) is used which leads to an approximately elastic limit prior to the onset of fracture. Displacement steps $\Delta u_y$ are applied on top of the domain on two straps $19.75\textrm{ mm} \leq x \leq 20.0\textrm{ mm}$ and $60.0\textrm{ mm} \leq x \leq 60.25\textrm{ mm}$ which are refined accordingly. On the bottom side of the specimen, the $y-$ and $z-$displacements are fixed at $x=4\textrm{ mm}$ and $x=76\textrm{ mm}$ while the $x-$displacements are set to zero only at $x=4\textrm{ mm}$. Coarse displacement steps of size $5\times10^{-3}\textrm{ mm}$ are applied up to crack nucleation when the step size is decreased to a value of $5\times10^{-4}\textrm{ mm}$. Following the discussion about the meaning of $l_0$ and its interpretation as a material parameter (\cite{tanne2018crack}), the length-scale parameter $l_0$ is chosen to reproduce the failure load for the $\gamma=0\degree$ specimen. In this case, a value $l_0 = 0.125$ mm was found to lead to satisfying results. The results for all three inclination angles using the same set of parameters are shown in \figref{failureloads} for ansatz orders $p=2$ and $p=3$. The load-displacement curves show the expected shape and with $p=3$ and a maximum refinement depth $k=5$ the phase-field seems to be sufficiently resolved. As observed in the experiments, with higher inclination angle the failure load increases. The phase-field model is able to capture this trend although the difference in loads is underestimated. The failure load is determined as the first local maximum of the load-displacement curve, which leads to a deviation of $0.33\%$ for the $0\degree$ specimen, a deviation of $8.85\%$ for the $30\degree$ specimen and a deviation of $10.55\%$ for the $45\degree$ case. For the latter two cases, the average of all experiments has been taken as reference value. \textcolor{changez}{The computations were done on a cluster node with 28 cores, 64 GB RAM and 2.6 GHz processor taking $57$h $58$min for the $0\degree$ case and $62$h $27$min for the $30\degree$ case ($p=3$)}.
\begin{table}[!h]
	\centering
	\setlength{\tabcolsep}{7pt}
	\renewcommand{\arraystretch}{1.2}
	\begin{tabular}{llccc} \toprule
		& &$0\degree$ & $30\degree$ & $45\degree$ \\ \midrule
		\multirow{ 2}{*}{$p=2$} &comp. load [N] & $0.70094$ & $0.78098$ & $0.9209$ \\ 
		&deviation [$\%$]& $4.69$ & $7.68$ & $10.4$ \\[1mm] 
		\multirow{ 2}{*}{$p=3$}& comp. load [N] & $0.67021$ & $0.77719$ & $0.91955 $ \\ 
		& deviation [$\%$] & $0.33$ & $8.13$ & $10.55$ \\ \bottomrule
	\end{tabular} \\[3.2mm]
	\caption{\textcolor{changez}{Computed failure loads and deviation from experiments in percentage}.}
	\label{tab:deviation}
\end{table}
\subsection{Fracture Surface}\label{sec:res-surface}
\begin{figure}[b!]
	\centering
	\includegraphics[width=0.99\textwidth]{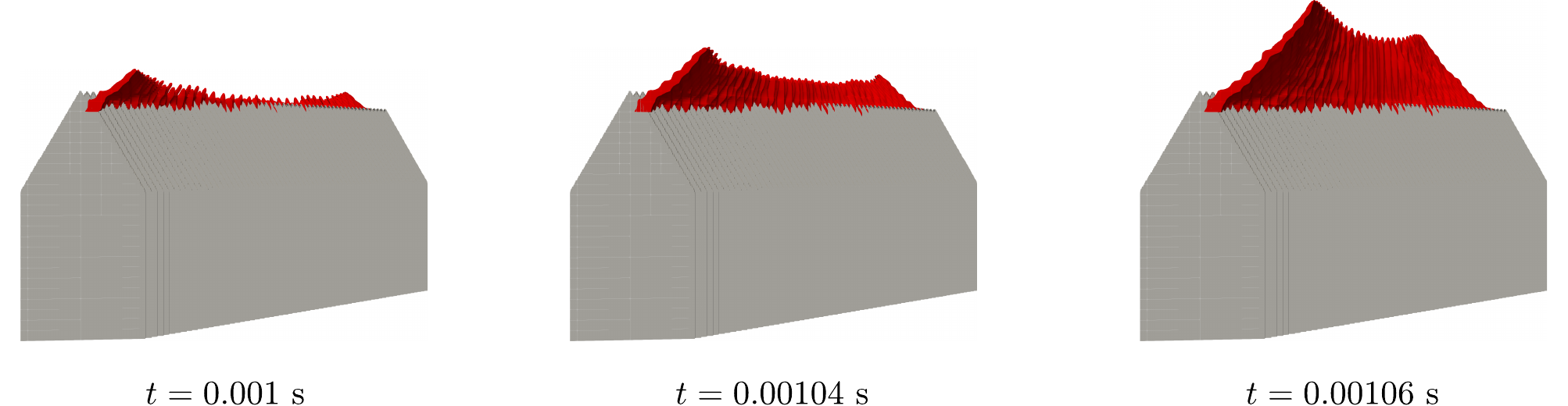}
	\caption{Initial evolution of the fracture surface for the $45 \degree$ specimen. To visualize the fracture an iso-volume for $s\leq0.03$ is extracted from the simulation results and shown in red.}
	\label{fig:surfaceEvolution}
\end{figure}
\begin{figure*}[!b]
	\centering
	\includegraphics[width=.95\textwidth]{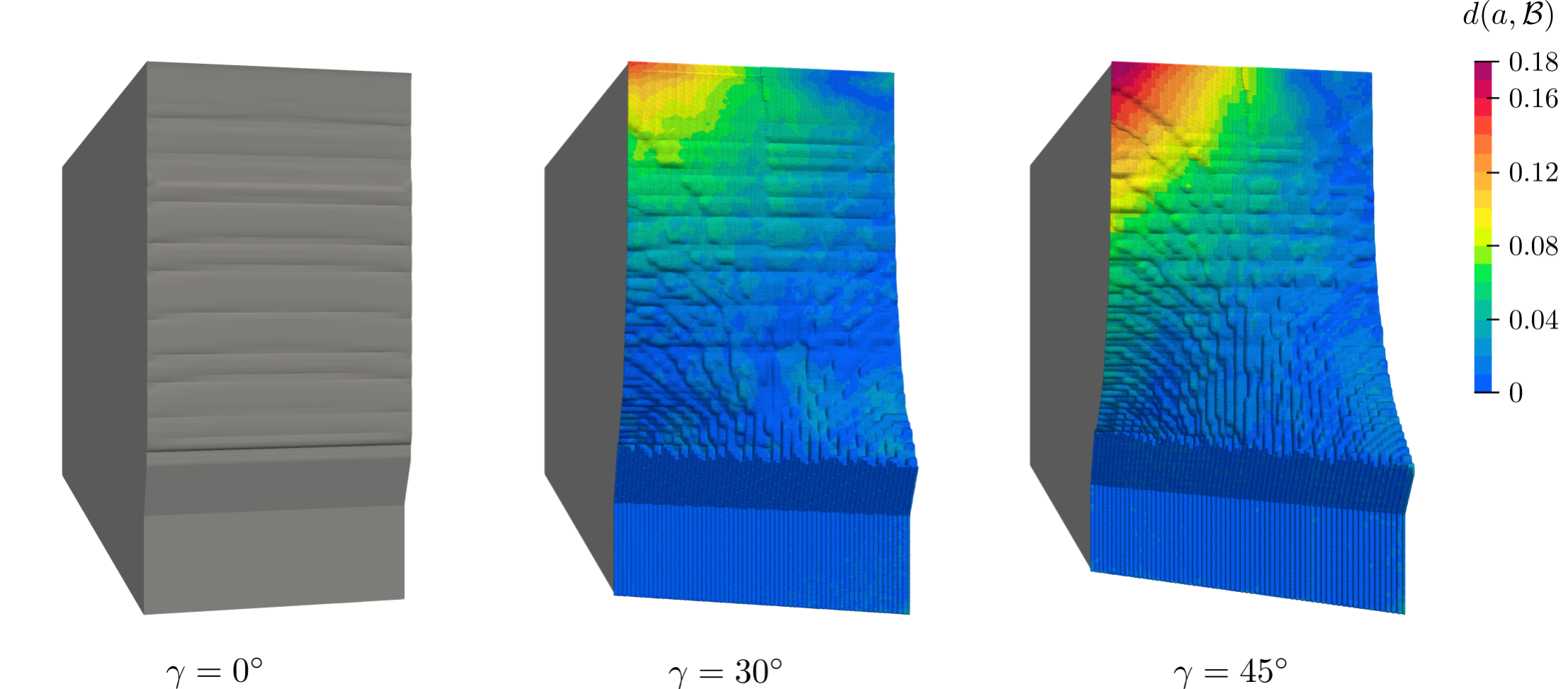}
	\caption{Distance to the reference point cloud evaluated on the \textcolor{changez}{computed} fracture surface for the $30\degree$ and $45\degree$ specimen. The distance is normalized by the length of the notch, which is $11.55$ mm for the $30\degree$ and $14.14$ mm for the $45\degree$ geometry  }
	\label{fig:surfaces}
\end{figure*}

As setup (a) results in unreasonable crack patterns when the dynamic model is used, different boundary conditions need to be applied. In contrast to the quasi-static setup the $y-$ and $z-$displacements are set to zero \textcolor{changez}{at the lower surface} on two straps $3.75 \leq x \leq 4.0$ and $76.0 \leq x \leq 76.25$, while $x-$displacements are not restrained on the bottom. For the dynamic simulation the computationally cheaper quadratic degradation function (Eq. \ref{eq:degradation}) is used as the crack path is independent of this choice. In a similar manner to the quasi-static simulation, coarse time steps of size $\Delta t = 1 \times 10^{-4}$ s are followed by finer time steps of size $\Delta t = 2 \times 10^{-5}$ s as soon as the crack initiates. A constant compressive load $\sigma=2$ Pa is applied on the two boundary straps on top of the specimen. A length-scale parameter $l_0 = 0.125$ with refinement depth $h=4$ and ansatz order $p=3$ results in a smooth fracture surface. In \cref{fig:surfaceEvolution}, initial crack growth is visualized for the $45 \degree$ geometry. Here, the crack has been extracted for three different time steps as an iso-volume for a phase-field value of $s\leq0.03$. As can be seen, crack nucleation starts towards the ends of the notch. In order to evaluate the distances to the point cloud crack surfaces need to be extracted from the numerical results. To this end, iso-volumes for $s\geq0.03$ are created resulting in two fractured parts of the specimen. In \text{Fig. \ref{fig:surfaces}}, the left half of the specimen, i.e. the one connected to corner $x=0.0$ mm, is visualized for each geometry. Here, the distances $d(a,\mathcal{B})$, where $a$ is a vertex in the numerical fracture surface and $\mathcal{B}$ is the set of points in the reference point cloud (\text{Eq. \ref{eq.dist-measure}}), are evaluated for the $30 \degree$ and $45 \degree$ geometry. For the $0 \degree$ case no specimens \textcolor{changez}{were available} to create a point cloud. Considering the results for the $30 \degree$ and $45 \degree$ specimen, one can clearly observe how crack initiation is captured in both cases. One effect which increases the error in the initiation region is non-physical widening of the crack path at later time steps when crack propagation speed decreases. This phenomenon is even more pronounced when a fully anisotropic formulation is used and might also be attributed to the introduction of the history variable which can lead to overestimation of the fracture surface energy \text{(\cite{gerasimov2018penalization})}. 
\begin{table}[!t]
	\centering
	\setlength{\tabcolsep}{7pt}
	\renewcommand{\arraystretch}{1.2}
	\begin{tabular}{lcccc} \toprule
		$\gamma$ [$\degree$] & $h_{\textrm{HD}}(\mathcal{A},\mathcal{B})$ [mm] & $h_{\textrm{HD}}(\mathcal{B},\mathcal{A})$ [mm] & $h_{\textrm{MHD}}(\mathcal{A},\mathcal{B})$ [mm] & $h_{\textrm{HD}}(\mathcal{B},\mathcal{A})$ [mm] \\ \midrule
		$30 $ & $1.731$& $ \boldsymbol{1.906}$ &$0.2101$ &$\boldsymbol{0.2388}$ \\
		$45 $ & $2.663$& $\boldsymbol{3.588}$ &$0.2751$ &$\boldsymbol{0.3588}$ \\ \bottomrule
	\end{tabular} \\[3.2mm]
	\caption{One-sided distance measures \cref{eq:hd} and \cref{eq:mhd} for the $30 \degree$ and $45 \degree$ geometry, where $\mathcal{A}$ is the set of vertices in the numerically obtained crack surface and $\mathcal{B}$ are the points in the reference point cloud.}
	\label{tab:hd}
\end{table}
In the upper part we see a deviation of the fracture path from the experimental one. While the phase-field model results in a fracture surface parallel to the $y$-$z$-plane, the experiment shows a clear curvature which is even more pronounced for the $45\degree$ specimen. Consequently, a higher deviation is observed in the latter case. It should be noted, that in one out of five experiments for the $30 \degree$ specimen an orthogonal crack as in the simulation could be observed. A different choice of tension-compression split could help to capture the curved crack path in the upper part of the specimen and needs further investigations. We conclude that the spectral split by \cite{miehe2010thermodynamically} shows clear deficits under the present loading conditions, as it moreover leads to saturation of phase-field propagation as soon as the crack propagates up to a height of $y=8$ mm if the quasi-static model is used. In \cref{tab:hd}, the one-sided Hausdorff and modified Hausdorff distances are listed for both geometries. The resulting Hausdorff distances $H_{\textrm{HD}}(\mathcal{A},\mathcal{B})$ and $H_{\textrm{MHD}}(\mathcal{A},\mathcal{B})$ are obtained following \cref{eq:H-hd} and \cref{eq:H-mhd} and correspond to the values marked in bold. For the computation of the distances only the front part of the specimen and, respectively, the point cloud has been considered, cf. \cref{fig:surfaces}. For the $30 \degree$ geometry a Hausdorff distance of $1.906$ mm and a modified (average) Hausdorff distance of $0.2388$ mm are obtained. Due to the more complex fracture surface higher deviations are obtained in the $45 \degree$ case resulting in a Hausdorff distance of $3.588$ mm and a modified Hausdorff distance of $0.3588$ mm.  

\begin{figure}[h!]
	\centering
	\includegraphics[width=0.92\textwidth]{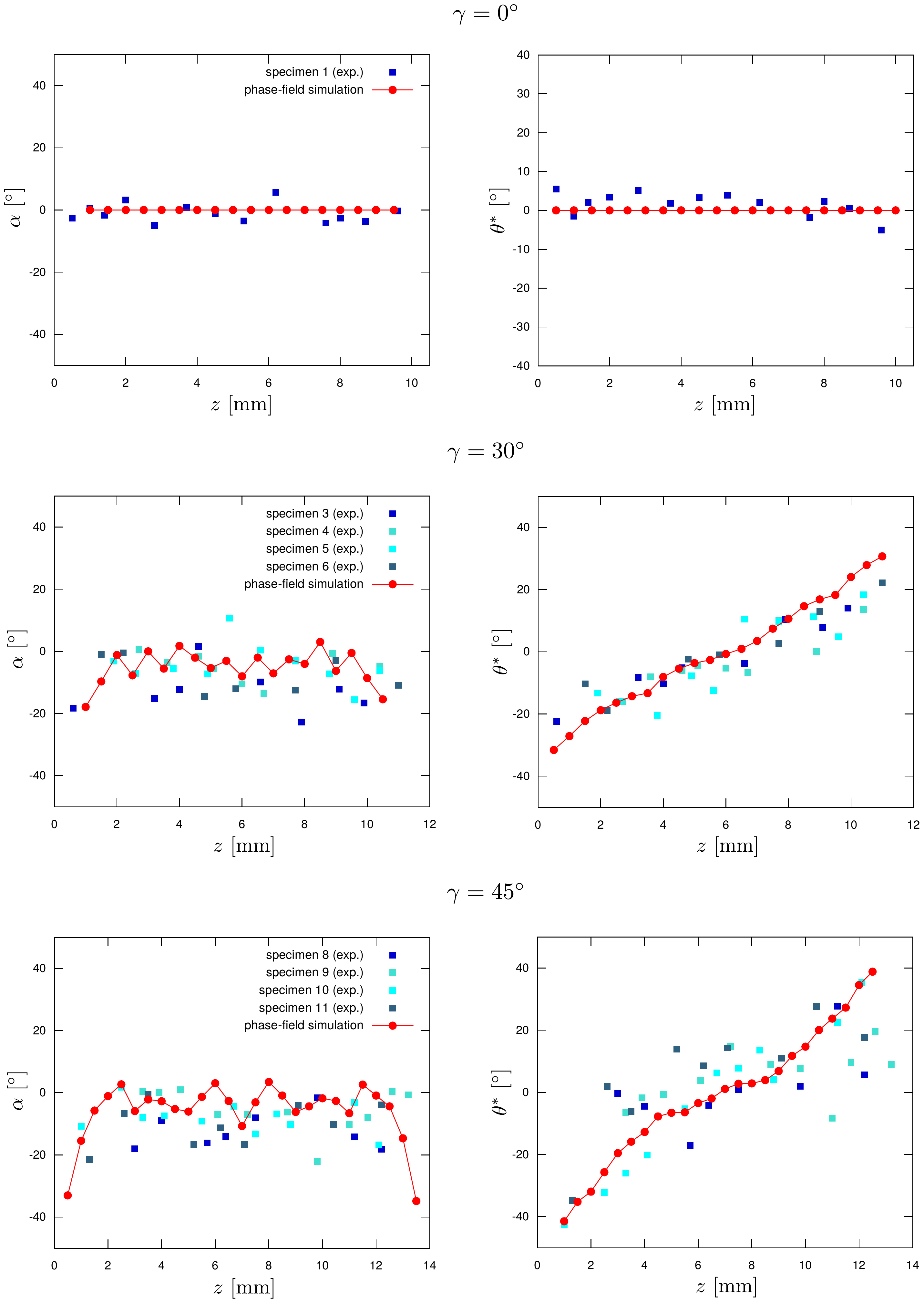}
	\caption{Comparison of computed \textcolor{changez}{crack initiation angles} $\alpha$ and $\theta^\ast$ angles with experimental results from \cite{yosibash:16}.}
	\label{fig:initiationAngles}
\end{figure}
\subsection{Initiation Angles}\label{sec:res-initangles}
\textcolor{changez}{To} determine the initiation angles, the \textcolor{changez}{computed} fracture surface up to a height of $y=6.5$ mm has been extracted and evaluated using a 3D modeling tool. The results for the three geometries are found along the experimental results by \cite{yosibash:16} in Figure \ref{fig:initiationAngles}, where the initiation angles are plotted over the $z$-axis. The initiation angle $\alpha$ has been determined as the approximated angle over a notch length of $\Delta z = 0.5$ mm, while $\theta^{\ast}$ is measured point-wise every $\Delta z = 0.5$ mm. 
For the $0 \degree$ specimen, experimental values for both initiation angles are approximately zero due to the straight crack surface nucleating directly on the notch tip for all $z$. This nucleation surface is clearly captured by the numerical \textcolor{changez}{simulation}. In the two other cases almost exclusively negative values for $\alpha$ are observed, which corresponds to an initiation plane rotated clockwise around the $y$-axis. This implies that the crack nucleates on the tip of the notch only in the center of the specimen ($z=5$ mm) shifting the initiation point in negative $x$-direction for increasing and in positive $x$-direction for decreasing $z$-values. The second inclination angle $\theta^{\ast}$ shows a point-symmetric trend developing from a negative angle $\theta^{\ast}_{30 \degree} =-31.5 \degree$ and $\theta^{\ast}_{45 \degree} =-45.5 \degree$ for $z=0$ mm to a similar absolute, however, positive angle toward the other end of the notch. This reflects the curvature of the fracture surface which becomes more pronounced with higher inclination angle. Comparing the results for the phase-field model with the experimental data further strengthens the observations in Section \ref{sec:res-surface} in the light of several experiments. The presented model is able to accurately capture nucleation of the fracture surface and lies within the range of experiments. 
 
 \section{Conclusions} \label{sec:conclusions}
In this contribution, a \textcolor{changez}{quantitative} benchmark problem for brittle fracture under mixed mode I + II + III loading has been presented. In the experiments, four-point bending tests \textcolor{changez}{were} conducted on graphite specimen\textcolor{changez}{s} containing a sharp V-notch at different inclination angles. Experimentally obtained failure loads and initiation angles along with 3D representations of the fractured specimen\textcolor{changez}{s} allow for a detailed \textcolor{changez}{and quantitative} comparison of fracture nucleation and propagation observed numerically.\\ 
The proposed benchmark is used to validate a phase-field model for brittle fracture based on FCM and multi-level hp-refinement. The phase-field model is
able to reproduce the geometry dependent failure loads although the difference in loads between the $0\degree$ and $30\degree$, as well as the $30\degree$ and $45\degree$ case are slightly underestimated. Here, the maximum deviation from experimental average is $\sim10 \%$ for all geometries. Evaluation of the fracture surface shows that the initial crack path in the lower half of the specimen can be reproduced almost exactly. This is confirmed by the geometric information measured in terms of two initiation angles, where the phase-field results lie almost exclusively within the range of experiments. The \textcolor{changez}{evolving} upper part of the crack surface deviates from the experiments. Here, the phase-field simulations predict a straight surface for the $30\degree$ and $45\degree$ geometries while the actual surface shows a curvature. This might be related to the choice of tension-compression split and requires further investigation.\\
All data is provided \textcolor{changes}{for download} along with the numerical results and a detailed description of evaluation processes to invite and enable validation of different numerical frameworks.

\section*{Acknowledgements} 
We gratefully acknowledge the support of the International School of Science and Engineering (IGSSE) \textcolor{changes}{of the Technical University of Munich} under project 12.03 \textcolor{changes}{'Fracture Risk in Human Bones'}.
\newpage
\bibliographystyle{apalike}
 \bibliography{library.bib}

\begin{thebibliography}{}

\bibitem[Alberti, 2000]{alberti2000variational}
Alberti, G. (2000).
\newblock Variational models for phase transitions, an approach via
  $\gamma$-convergence.
\newblock In {\em Calculus of variations and partial differential equations},
  pages 95--114. Springer.

\bibitem[Ambati et~al., 2015]{ambati2015review}
Ambati, M., Gerasimov, T., and Lorenzis, L.~D. (2015).
\newblock A review on phase-field models of brittle fracture and a new fast
  hybrid formulation.
\newblock {\em Computational Mechanics}, 55(2):383--405.

\bibitem[Amor et~al., 2009]{amor2009regularized}
Amor, H., Marigo, J.-J., and Maurini, C. (2009).
\newblock Regularized formulation of the variational brittle fracture with
  unilateral contact: Numerical experiments.
\newblock 57:1209--1229.

\bibitem[Babuska et~al., 1981]{babuska1981p}
Babuska, I., Szabo, B.~A., and Katz, I.~N. (1981).
\newblock The p-version of the finite element method.
\newblock {\em SIAM journal on numerical analysis}, 18(3):515--545.

\bibitem[Belytschko and Black, 1999]{bely1999elastic}
Belytschko, T. and Black, T. (1999).
\newblock Elastic crack growth in finite elements with minimal remeshing.
\newblock {\em International journal for numerical methods in engineering},
  45(5):601--620.

\bibitem[Borden, 2012]{borden2012isogeometric}
Borden, M.~J. (2012).
\newblock {\em Isogeometric analysis of phase-field models for dynamic brittle
  and ductile fracture}.
\newblock PhD thesis.

\bibitem[Borden et~al., 2014]{borden2014higher}
Borden, M.~J., Hughes, T.~J., Landis, C.~M., and Verhoosel, C.~V. (2014).
\newblock A higher-order phase-field model for brittle fracture: Formulation
  and analysis within the isogeometric analysis framework.
\newblock {\em Computer Methods in Applied Mechanics and Engineering},
  273:100--118.

\bibitem[Borden et~al., 2012]{borden2012phase}
Borden, M.~J., Verhoosel, C.~V., Scott, M.~A., Hughes, T.~J., and Landis, C.~M.
  (2012).
\newblock A phase-field description of dynamic brittle fracture.
\newblock {\em Computer Methods in Applied Mechanics and Engineering},
  217:77--95.

\bibitem[Bourdin et~al., 2000]{bourd2000}
Bourdin, B., Francfort, G.~A., and Marigo, J.-J. (2000).
\newblock Numerical experiments in revisited brittle fracture.
\newblock {\em Journal of the Mechanics and Physics of Solids}, 48(4):797--826.

\bibitem[Bourdin et~al., 2008]{bour2008}
Bourdin, B., Francfort, G.~A., and Marigo, J.-J. (2008).
\newblock The variational approach to fracture.
\newblock {\em Journal of elasticity}, 91(1-3):5--148.

\bibitem[Carpiuc-Prisacari et~al., 2017]{carpiuc2017complex}
Carpiuc-Prisacari, A., Poncelet, M., Kazymyrenko, K., Leclerc, H., and Hild, F.
  (2017).
\newblock A complex mixed-mode crack propagation test performed with a 6-axis
  testing machine and full-field measurements.
\newblock {\em Engineering Fracture Mechanics}, 176:1--22.

\bibitem[Citarella and Buchholz, 2008]{citarella2008comparison}
Citarella, R. and Buchholz, F.-G. (2008).
\newblock Comparison of crack growth simulation by dbem and fem for
  sen-specimens undergoing torsion or bending loading.
\newblock {\em Engineering Fracture Mechanics}, 75(3-4):489--509.

\bibitem[Cloud{C}ompare, 2019]{cloudcomp}
Cloud{C}ompare (2019).
\newblock Version 2.10.1.
\newblock \url{http://www.cloudcompare.org}.

\bibitem[Dally and Weinberg, 2017]{dally2017phase}
Dally, T. and Weinberg, K. (2017).
\newblock The phase-field approach as a tool for experimental validations in
  fracture mechanics.
\newblock {\em Continuum Mechanics and Thermodynamics}, 29(4):947--956.

\bibitem[Di~Stolfo et~al., 2016]{di2016easy}
Di~Stolfo, P., Schr{\"o}der, A., Zander, N., and Kollmannsberger, S. (2016).
\newblock An easy treatment of hanging nodes in hp-finite elements.
\newblock {\em Finite Elements in Analysis and Design}, 121:101--117.

\bibitem[Dubuisson and Jain, 1994]{dubuisson1994modified}
Dubuisson, M.-P. and Jain, A.~K. (1994).
\newblock A modified hausdorff distance for object matching.
\newblock In {\em Proceedings of 12th international conference on pattern
  recognition}, volume~1, pages 566--568. IEEE.

\bibitem[D{\"u}ster et~al., 2008]{duster2008finite}
D{\"u}ster, A., Parvizian, J., Yang, Z., and Rank, E. (2008).
\newblock The finite cell method for three-dimensional problems of solid
  mechanics.
\newblock {\em Computer methods in applied mechanics and engineering},
  197(45-48):3768--3782.

\bibitem[Francfort and Marigo, 1998]{fran1998revisiting}
Francfort, G.~A. and Marigo, J.-J. (1998).
\newblock Revisiting brittle fracture as an energy minimization problem.
\newblock {\em Journal of the Mechanics and Physics of Solids},
  46(8):1319--1342.

\bibitem[G{\'a}lvez et~al., 1998]{galvez1998mixed}
G{\'a}lvez, J., Elices, M., Guinea, G., and Planas, J. (1998).
\newblock Mixed mode fracture of concrete under proportional and
  nonproportional loading.
\newblock {\em International Journal of Fracture}, 94(3):267--284.

\bibitem[Gerasimov and De~Lorenzis, 2016]{gerasimov2016line}
Gerasimov, T. and De~Lorenzis, L. (2016).
\newblock A line search assisted monolithic approach for phase-field computing
  of brittle fracture.
\newblock {\em Computer Methods in Applied Mechanics and Engineering},
  312:276--303.

\bibitem[Gerasimov and De~Lorenzis, 2018]{gerasimov2018penalization}
Gerasimov, T. and De~Lorenzis, L. (2018).
\newblock On penalization in variational phase-field models of brittle
  fracture.
\newblock {\em arXiv preprint arXiv:1811.05334}.

\bibitem[Griffith and Eng, 1921]{griffith1921}
Griffith, A.~A. and Eng, M. (1921).
\newblock Vi. the phenomena of rupture and flow in solids.
\newblock {\em Phil. Trans. R. Soc. Lond. A}, 221(582-593):163--198.

\bibitem[Hennig et~al., 2016]{hennig2016bezier}
Hennig, P., M{\"u}ller, S., and K{\"a}stner, M. (2016).
\newblock B{\'e}zier extraction and adaptive refinement of truncated
  hierarchical nurbs.
\newblock {\em Computer Methods in Applied Mechanics and Engineering},
  305:316--339.

\bibitem[Hesch et~al., 2016]{hesch2016isogeometric}
Hesch, C., Schu{\ss}, S., Dittmann, M., Franke, M., and Weinberg, K. (2016).
\newblock Isogeometric analysis and hierarchical refinement for higher-order
  phase-field models.
\newblock {\em Computer Methods in Applied Mechanics and Engineering},
  303:185--207.

\bibitem[Hubrich et~al., 2017]{hubrich2017numerical}
Hubrich, S., Di~Stolfo, P., Kudela, L., Kollmannsberger, S., Rank, E.,
  Schr{\"o}der, A., and D{\"u}ster, A. (2017).
\newblock Numerical integration of discontinuous functions: moment fitting and
  smart octree.
\newblock {\em Computational Mechanics}, 60(5):863--881.

\bibitem[Irwin, 1958]{Irwi1958elasticity}
Irwin, G.~R. (1958).
\newblock {\em Fracture}, pages 551--590.
\newblock Springer Berlin Heidelberg, Berlin, Heidelberg.

\bibitem[Jeong et~al., 2018]{jeong2018phase}
Jeong, H., Signetti, S., Han, T.-S., and Ryu, S. (2018).
\newblock Phase field modeling of crack propagation under combined shear and
  tensile loading with hybrid formulation.
\newblock {\em Computational Materials Science}, 155:483--492.

\bibitem[Kudela et~al., 2018]{kudela2018image}
Kudela, L., Frischmann, F., Yossef, O.~E., Kollmannsberger, S., Yosibash, Z.,
  and Rank, E. (2018).
\newblock Image-based mesh generation of tubular geometries under circular
  motion in refractive environments.
\newblock {\em Machine Vision and Applications}, 29(5):719--733.

\bibitem[Miehe et~al., 2010a]{miehe2010phase}
Miehe, C., Hofacker, M., and Welschinger, F. (2010a).
\newblock A phase field model for rate-independent crack propagation: Robust
  algorithmic implementation based on operator splits.
\newblock {\em Computer Methods in Applied Mechanics and Engineering},
  199(45-48):2765--2778.

\bibitem[Miehe et~al., 2010b]{miehe2010thermodynamically}
Miehe, C., Welschinger, F., and Hofacker, M. (2010b).
\newblock Thermodynamically consistent phase-field models of fracture:
  Variational principles and multi-field fe implementations.
\newblock {\em International Journal for Numerical Methods in Engineering},
  83(10):1273--1311.

\bibitem[Mo{\"e}s et~al., 1999]{moes1999finite}
Mo{\"e}s, N., Dolbow, J., and Belytschko, T. (1999).
\newblock A finite element method for crack growth without remeshing.
\newblock {\em International journal for numerical methods in engineering},
  46(1):131--150.

\bibitem[Nagaraja et~al., 2018]{nagaraja:18}
Nagaraja, S., Elhaddad, M., Ambati, M., Kollmannsberger, S., De~Lorenzis, L.,
  and Rank, E. (2018).
\newblock Phase-field modeling of brittle fracture with multi-level hp-fem and
  the finite cell method.
\newblock {\em arXiv preprint arXiv:1804.08380}.

\bibitem[Nguyen et~al., 2016]{nguyen2016initiation}
Nguyen, T.~T., Yvonnet, J., Bornert, M., and Chateau, C. (2016).
\newblock Initiation and propagation of complex 3d networks of cracks in
  heterogeneous quasi-brittle materials: Direct comparison between in situ
  testing-microct experiments and phase field simulations.
\newblock {\em Journal of the Mechanics and Physics of Solids}, 95:320--350.

\bibitem[Nguyen et~al., 2015]{nguyen2015phase}
Nguyen, T.~T., Yvonnet, J., Zhu, Q.-Z., Bornert, M., and Chateau, C. (2015).
\newblock A phase field method to simulate crack nucleation and propagation in
  strongly heterogeneous materials from direct imaging of their microstructure.
\newblock {\em Engineering Fracture Mechanics}, 139:18--39.

\bibitem[Nooru-Mohamed et~al., 1993]{nooru1993experimental}
Nooru-Mohamed, M., Schlangen, E., and van Mier, J.~G. (1993).
\newblock Experimental and numerical study on the behavior of concrete
  subjected to biaxial tension and shear.
\newblock {\em Advanced cement based materials}, 1(1):22--37.

\bibitem[Nooru-Mohamed, 1993]{nooru1993mixed}
Nooru-Mohamed, M.~B. (1993).
\newblock Mixed-mode fracture of concrete: An experimental approach.

\bibitem[Ortiz and Pandolfi, 1999]{ortiz1999finite}
Ortiz, M. and Pandolfi, A. (1999).
\newblock Finite-deformation irreversible cohesive elements for
  three-dimensional crack-propagation analysis.
\newblock {\em International journal for numerical methods in engineering},
  44(9):1267--1282.

\bibitem[Parvizian et~al., 2007]{parvizian2007finite}
Parvizian, J., D{\"u}ster, A., and Rank, E. (2007).
\newblock Finite cell method.
\newblock {\em Computational Mechanics}, 41(1):121--133.

\bibitem[R{\'e}thor{\'e}, 2018]{rethore2018PMMA}
R{\'e}thor{\'e}, J. (2018).
\newblock P{M}{M}{A} {M}ixed mode fracture ({V}ersion 1.0) [{D}ata set].
  {Z}enodo.
\newblock \url{http://doi.org/10.5281/zenodo.1473126}.

\bibitem[Schillinger et~al., 2012]{schillinger2012small}
Schillinger, D., Ruess, M., Zander, N., Bazilevs, Y., D{\"u}ster, A., and Rank,
  E. (2012).
\newblock Small and large deformation analysis with the p-and b-spline versions
  of the finite cell method.
\newblock {\em Computational Mechanics}, 50(4):445--478.

\bibitem[Shao et~al., 2019]{shao2019adaptive}
Shao, Y., Duan, Q., and Qiu, S. (2019).
\newblock Adaptive consistent element-free galerkin method for phase-field
  model of brittle fracture.
\newblock {\em Computational Mechanics}, pages 1--27.

\bibitem[Tann{\'e} et~al., 2018]{tanne2018crack}
Tann{\'e}, E., Li, T., Bourdin, B., Marigo, J.-J., and Maurini, C. (2018).
\newblock Crack nucleation in variational phase-field models of brittle
  fracture.
\newblock {\em Journal of the Mechanics and Physics of Solids}, 110:80--99.

\bibitem[Triggs et~al., 1999]{triggs1999bundle}
Triggs, B., McLauchlan, P.~F., Hartley, R.~I., and Fitzgibbon, A.~W. (1999).
\newblock Bundle adjustment—a modern synthesis.
\newblock In {\em International workshop on vision algorithms}, pages 298--372.
  Springer.

\bibitem[Yosibash and Mittelman, 2016]{yosibash:16}
Yosibash, Z. and Mittelman, B. (2016).
\newblock A 3-d failure initiation criterion from a sharp v-notch edge in
  elastic brittle structures.
\newblock {\em European Journal of Mechanics-A/Solids}, 60:70--94.

\bibitem[Zander et~al., 2015]{zander2015multi}
Zander, N., Bog, T., Kollmannsberger, S., Schillinger, D., and Rank, E. (2015).
\newblock Multi-level hp-adaptivity: high-order mesh adaptivity without the
  difficulties of constraining hanging nodes.
\newblock {\em Computational Mechanics}, 55(3):499--517.

\end{thebibliography}

\end{document}